# Humoral immunocompetence in relation to condition, size, asymmetry and MHC class II variation in great snipe (*Gallinago media*) males

R. EKBLOM[1]*, D. HASSELQUIST[2], S. A. SÆTHER[1, 3], P. FISKE[4], J. A. KÅLÅS[4], M. GRAHN[5] and J. HÖGLUND[1]

1. Population Biology/Evolutionary Biology Centre, Uppsala University, Norbyv.18D, SE-752 36 Uppsala, Sweden
2. Department of Animal Ecology, Lund University, Ecology Building, SE-223 62 Lund, Sweden
3. Evolutionary Biology/Evolutionary Biology Centre, Uppsala University, Norbyv.18D, SE-752 36 Uppsala, Sweden
4. Norwegian Institute for Nature Research, Tungasletta 2, N-7485, Trondheim, Norway
5. Södertörn University College, Box 4101, SE-141 04 Huddinge, Sweden

*e-mail: robert.ekblom@ebc.uu.se

*Short title: Immunocompetence in great snipe*

**Summary**

**1.** In recent years many studies have investigated the relationships between different aspects of the immune system and ecology in various organisms. Yet, it remains unclear why individuals differ in their ability to mount an immune response against various antigens (often referred to as "immunocompetence"). Different kinds of trade-offs may be involved and costs of mounting the immune response often lead to condition dependent effects.
**2.** We investigated how variation in condition, morphology and genetic variables influenced the amount of antibodies produced against two novel antigens in a migrating bird, the great snipe (*Gallinago media*).
**3.** We found no evidence for condition dependence of the antibody response and no effect of MHC genetics. There was, however, a weak negative correlation between body size and the amount of antibody production, which may indicate a trade-off between growth and immune response in this species.

# Introduction

During the last decade much research has been conducted in the field of "immunoecology" (Sheldon & Verhulst 1996; Westneat & Birkhead 1998; Owens & Wilson 1999). The immune system of both vertebrates (Norris & Evans 2000) and invertebrates (Ryder & Siva-Jothy 2000; Cotter, Kruuk & Wilson 2004) has been used to address important ecological and evolutionary issues such as life history decisions (Norris & Evans 2000; Cotter *et al.* 2004), sexual selection (Hamilton & Zuk 1982; Folstad & Karter 1992; Kurtz & Sauer 1999) and co-evolution between hosts and parasites (Haldane 1949; Dybdahl & Lively 1998). Still, we do not really understand exactly what determines how strong reaction an individual will produce against a certain antigen.

The vertebrate immune system can broadly be divided into innate and acquired responses. The former is responsible for inflammation and phagocytosis. These reactions constitute a first line of defence against invading parasites and have much in common with immune response of invertebrates. The acquired immune response has the property of improving over time through the process of learning and is highly specific. Because of this immunological memory, the immune system is able to mount a stronger response to subsequent repeated infections, a so-called secondary response. There are two major branches of the acquired response: cellular and humoral immunity. The cellular branch is regulated by T lymphocytes (TH1) and acts mainly against intra-cellular microbes and viruses. The main actors in the humoral response are T lymphocytes (TH2) and B lymphocytes (which produce antibodies), and this branch primarily protects against inter-cellular pathogens and parasites. For a more extensive review of the immune system see for example Roitt (1997).

Agents that have the ability to inflict an immune reaction are called antigens. The ability of an individual to mount an immune response to a certain antigen, either a pathogen or a harmless substance, is often referred to as "immunocompetence" (Owens & Wilson 1999). Mounting such immune response may be energetically costly (Ots *et al.* 2001; Martin II, Scheuerlein & Wikelski 2003, but see Svensson *et al.* 1998), and must therefore be traded off against other important traits such as growth and expression of sexually selected characters (Lochmiller & Deerenberg 2000). This has the consequence that the ability to mount an immune response is often influenced by the condition of the individual (Coltman *et al.* 2001). There are also direct costs associated with mounting the immuneresponse. For example an immune response leads to formation of free radicals that may damage host tissues (von Schantz *et al.* 1999). The immune system may also attack the bodies' own tissues causing immunopathology such as autoimmune diseases (Råberg *et al.* 1998). Such physiological costs associated with mount-ing an immune response may ultimately even impair survival, as shown in a study on wild common eiders (*Somateria mollisima*; Hanssen *et al.* 2004).

Many studies have shown that there is a heritable component mediating the strength of the immune response both in vertebrates (e.g., Coltman *et al.* 2001; Råberg, Stjernman & Hasselquist 2003) and in invertebrates (Huang, Higgins & Buschman 1999; Kraaijeveld & Godfray 1999; Cotter *et al.* 2004). These studies have not addressed the question of which genes are responsible for this effect, but genes coding for proteins involved in the process of mounting an immune response seem to be good candidates. The proteins encoded by the MHC (Major Histocompatibility Complex) genes are responsible for self, non-self discrimination and triggering of the acquired immune system. These are among the most polymorphic genes known and there may be several hundred alleles at a certain locus (Edwards, Nusser & Gasper 2000; Bernatchez & Landry 2003). The different MHC alleles present their own set of antigens to the immune system and an individuals' MHC haplotype thus restricts which antigens are able to trigger the immune response (Suri *et al.* 2003). These properties make the MHC a very good candidate for a genetic link to the strength of the immune response. Indeed, it has been shown that selection for antibody response results in a change in MHC genotype frequencies (Pinard *et al.* 1993a-b; Dunnington *et al.* 1996). It is also well established, both in fish (Langefors *et al.* 2001; Lohm *et al.* 2002), mammals (Carrington *et al.* 1999; Jeffery *et al.* 2000; Penn, Damjanovich & Potts 2002) and birds (Bacon 1987; Lamont 1991), that MHC genotype influence resistance to many different kinds of diseases and parasites (reviewed in Jeffery & Bangham 2000).

Instead of studying the immune system by its effects on parasites, it may be advantageous to investigate immunocompetence directly by infecting the organism with a harmless antigen and follow the immune response (Sheldon & Verhulst 1996). Such "immune challenge tests" are now standard methods in immunoecology (Hasselquist *et al.* 1999; Hasselquist, Wasson & Winkler 2001), and have revealed a link between MHC genetics and immune response in chicken (Liu, Miller & Lamont 2002; Zhou & Lamont 2003). The aim of the present study was to investigate the influence of condition dependent factors and MHC variation on the strength of the humoral immune response in the great snipe (*Gallinago media*). We used two different harmless antigens (diphtheria and tetanus combined in a vaccine for normal use in humans) and followed specific antibody production to each of these antigens. Furthermore we studied the MHC genetics of great snipe by investigating the entire second exon of the MHC class II B gene (Ekblom, Grahn & Höglund 2003).

**Materials and methods**

FIELD STUDY

The great snipe is a migrating wader species, breeding in northern Europe and wintering in Africa. Our study site is located in the Dovre mountains in central Norway (for a general description of the species and the field site see Løfaldli, Kålås & Fiske 1992; Fiske & Kålås 1995). Male birds display in groups on leks from mid May to the beginning of July. We captured birds with mist-nets on the leks in the end of May and beginning of June in the years 2001 and 2002. Each captured male was ringed and we also collected blood and morphological data before releasing it. See Höglund *et al.* (1990) for information on methods used for measuring morphometry. Data from 51 males is included in this study.

MEASUREMENT OF THE ANTIBODY RESPONSE

We measured the amount of antibodies to two novel antigens (diphtheria and tetanus toxoid) by injecting males on the leks with 100 μl of diphtheria and tetanus vaccine. Before the vaccination, a control blood sample was taken from each bird. After approximately 12 days the birds were re-caught and a response blood sample was taken (the number of days varied from 11 to 15 but 45 out of 51 males were re-caught after exactly 12 days). Plasma was extracted from the blood after centrifugation and the amount of antigen-specific antibodies in the plasma was measured using an enzyme-linked immunosorbent assay (ELISA) procedure. For a more detailed description of the immunisation and immunoassay see Hasselquist *et al.* (2001, 1999) and Ekblom *et al.* (in press B). All antibody concentrations was measured as the slope of the substrate conversion over time measured in the units $10^{-3} \times$ optical density per minute (mOD/min) (measured on a $V_{max}$ ELISA plate reader, Molecular Devises, Sunnyvale CA, and analyzed using KineticCalc software, Winooski), with a higher slope indicating a higher concentration of specific antibodies in the sample. Control and response samples were run in duplicate on the ELISA reader and the average of these were used as antibody titers in all analyses. Antibody response was defined as antibody titer in the response sample minus the antibody titer in the control sample. The levels of antibodies in the control samples differed between the two years and the responses in 2002 were corrected for these differences (see Ekblom et al in press B).

GENETIC ANALYSES

Blood was sampled from the wing vein and stored in salt saturated DMSO. We extracted DNA using a standard phenol-chloroform method (Sambrook, Fritsch and Maniatis 1989) or with a DNeasy Tissue Kit (Qiagen). Extracted DNA was stored in water at -20°C. We typed the whole of the second exon of MHC class II B genes by separating PCR products, produced by using each of the primer-pairs Gint 1A-Int 2CGC and Gint 1A-Int 2AGC, with different sequences using a combination of DGGE and CDGE (Wu *et al.* 1999). Bands were than cut out and sequenced separately (see Ekblom *et al.* in press A, for a description of the molecular methods). Sequences were analysed using BioEdit 5·0·9 (Hall 1999) and MEGA 2·1 (Kumar *et al.* 2001). In the 51 males used in this study a total of 21 alleles were found (nucleotide sequences are deposited in GeneBank with accession numbers AF 485413-AF485417, AY620003-AY620004, AY620006-AY620018 and AY620021). Each individual had one to three different alleles, this may be because of a recent duplication affecting some of the alleles or because of amplification of alleles from more than one locus. In order to facilitate analysis, the alleles were grouped into six allelic lineages according to their nucleotide similarity (table 1; Ekblom *et al.* in press A).

**Table 1.** Number of nucleotide differences within and between the six different allelic lineages of MHC class II B sequences used in this study. Standard Error (within parentheses) was calculated from 1000 bootstrap replications.

| Name of allelic lineage | Number of alleles included | Mean number of nucleiotide differences within lineage | Mean number of nucleotide differences between lineages |
|---|---|---|---|
| *a* | 2 | 1·00 (0·93) | 13·13 (3·16) |
| *b* | 3 | 5·33 (1·72) | 11·83 (2·75) |
| *c* | 4 | 3·67 (1·32) | 10·03 (2·53) |
| *d* | 5 | 2·80 (1·20) | 9·04 (2·36) |
| *e* | 2 | 6·00 (2·33) | 10·14 (2·51) |
| *f* | 5 | 7·00 (1·81) | 12·75 (2·81) |

STATISTICAL ANALYSES

Because the distributions of the antigen responses were highly skewed the values were log transformed before analyses. All probability values are two-tailed. For statistical analyses we used SPSS 11·5.

VARIABLES IN THE REGRESSION ANALYSIS

*Antibody response*: We combined the variables for the antibody response to diphtheria and tetanus vaccine using a principal component analysis (PCA) (see Kilpimaa, Alatalo & Siitari 2004). The first principal component explained 70·4% of the variance and was used as a measure of total antibody response in further analyses. The values of this variable did not differ from a normal distribution (Kolmogorov-Smirnov test, Z = 0·726, n = 51, p = 0·67).

*Size*: We calculated body size as the first principal component of 3 different morphological measurements (tarsus length, total head length and wing length). This variable explained 61·6% of the variance of the three measurements.

*Condition*: Condition was calculated as the residuals of a regression between body mass and size (Brown 1996) controlled for time of the night and date, both of which are known to be correlated with body mass (Höglund, Kålås & Löfaldi 1990). This regression was highly significant ($F_{3,50}$ = 16·97, p < 0·001).

*Asymmetry*: As a measurement of body asymmetry we used the first principal component of the absolute difference of left side and right side values of three different two sided measurements (tarsus length, wing length, and the amount of white on tail of the outermost tail feather; Fiske, Kålås & Sæther 1994). This variable explained 48·8% of the total absolute differences between these traits.

*Whiteness on the tail*: As a measure of the amount of white on the tail (a trait possibly subjected to sexual selection; Höglund, Eriksson & Lindell 1990; Sæther *et al.* 2000), we used the mean of the length of the white area on the outermost tail feather on the left and right side of the tail, from the feather tip to the first dark spot (Sæther *et al.* 2000, Fiske *et al.* 1994).

*MHC heterozygosity*: Heterozygosity in the MHC class II B gene was approximated as the number of alleles (one to three) found in each individual using the PCR based typing system (see Genetic Analyses).

*Age*: In this species it is possible to determine whether a caught bird is one year old or older by determining the feather wear (Sæther, Kålås & Fiske 1994). We used minimum estimate of age, where an old bird caught for the first time was assumed to be two years old.

*Days to resampling*: The number of days between immunisation and resampling was also included in the analysis.

**Results**

RESPONSE TO IMMUNISATION

In 2001, we immunized 59 males and 33 of these were re-caught. In 2002 the corresponding figures were 24 re-caught out of 47 vaccinated. Birds responded to the vaccination by producing specific antibodies to each of the two antigens (Paired t-test, tetanus, t = -11·943, df = 50, p < 0·001; diphtheria, t = -4·625, df = 50, p < 0·001). Six individuals were immunized and re-caught in both years of the study, these did not produce a higher antibody titer after the second vaccination as compared to their primary immune response (Paired t-test, tetanus, t = -0·142, df = 5, p = 0·893; diphtheria, t = -1·712, df = 5, p = 0·148; Unpaired t-test, tetanus, t = 0·938, df = 55, p = 0·352; diphtheria, t = -0·027, df = 55, p = 0·978, Fig. 1). The responses of these six individuals tended to be positively correlated for the two study years, even though this was not significant, probably due to the low sample size (tetanus, r = 0·792, n = 6, p = 0·060; diphtheria, r = 0·601, n = 6, p = 0·207). Only the primary immune response from each individual was used in further analyses.

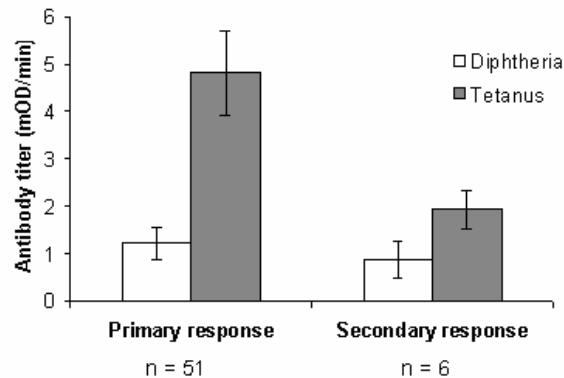

**Fig. 1.** Strength of the immune response (± SE), as measured by the difference in antibody titer between control sample and response sample. The response to tetanus is higher than the response to diphtheria. The response to a secondary injection was not stronger than to the primary injection.

The response against diphtheria and tetanus antigen in each individual was highly correlated (r = 0·408, n = 51, p = 0·003, Fig. 2) and the response was stronger to tetanus than to diphtheria (Paired t-test, t = -7·233, df = 50, p < 0·001, Fig. 2). This relationship appeared to be non-liear and individuals that responded strongly to diphtheria almost always had a strong tetanus response whereas individuals with a strong tetanus response could have any kind of response against diphteria (see Fig. 2). In further analyses, we used a combination of the two responses (from a PCA, see methods). There were no differences in antibody response between the two years of this study (t = -1·041, df = 49, p = 0·303).

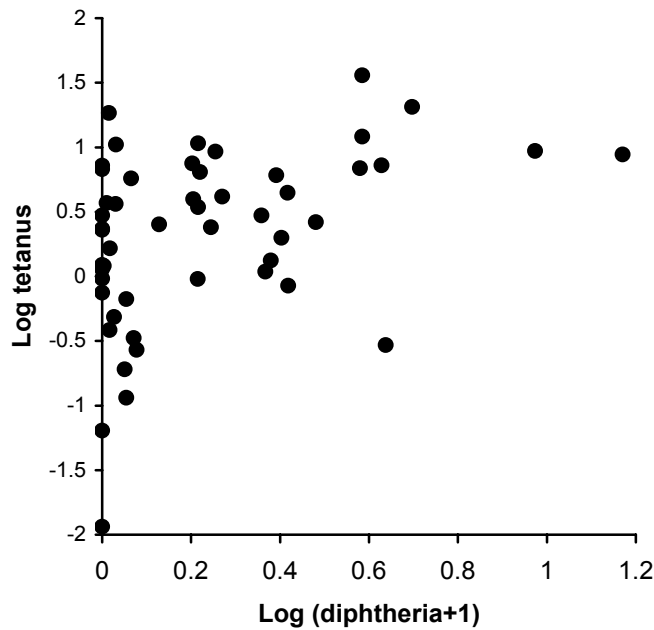

**Fig. 2.** Relationship between antibody responses to diphtheria and tetanus antigens (for statistics see text).

IMMUNE RESPONSE AND MHC TYPE

We did not find any differences between the allelic lineages in the amount of antibody production (ANOVA, $F_{5,109}$ = 1·10, p = 0·36). We also made another type of analysis, grouping antibody response into either high response, defined as one or both of the two responses being higher than 1·5 mOD/min, or low response, if both were lower than 1·5 mOD/min. Still there was no difference in how many times alleles of the different lineages were present in individuals from the two groups (Fisher's exact test, p = 0·64). We also ana-lysed whether there was any effect on the antibody response that could be caused by the combination of allelic lineages (genotype) in an individual. Again, no such effect was found (ANOVA, $F_{19,50} = 0.90$ p = 0.59) and there was no effect on the antibody response of the number of MHC alleles (heterozygosity) (ANOVA, $F_{2,48} = 0.94$ p = 0.40).

**Table 2.** Linear regression of the humoral immune response (antibody titer) in the great snipe.

| Variable | B | Std Error | Standardised Beta | t | p |
| --- | --- | --- | --- | --- | --- |
| (Intercept) | 2·82 | 3·07 | | 0·92 | 0·36 |
| Size | -0·37 | 0·16 | -0·37 | -2·34 | 0·024 |
| Condition | 0·005 | 0·030 | 0·027 | 0·16 | 0·87 |
| Asymmetry | 0·032 | 0·15 | 0·032 | 0·21 | 0·84 |
| White tail | -0·007 | 0·054 | -0·021 | -0·13 | 0·90 |
| MHC heterozygosity | 0·18 | 0·28 | 0·098 | 0·66 | 0·52 |
| Age | 0·068 | 0·13 | 0·081 | 0·50 | 0·62 |
| Days to re-sampling | -0·27 | 0·24 | -0·17 | -1·13 | 0·27 |

CONDITION DEPENDENCE OF THE IMMUNE RESPONSE

We analysed possible effects of condition-dependent traits and MHC heterozygosity on the antibody response, using a linear regression. The only variable with an significant effect was size (Table 2) and the full model was not significant ($R^2 = 0.14$, $F_{7,43} = 0.99$, p = 0.45). A model with size as the only variable was significant ($R^2 = 0.10$, $F_{1,49} = 5.44$, p = 0.024, Fig. 3). There was a weak negative correlation between size and immune response (r = -0.32, n = 51, p = 0.024, Fig. 3). This correlation was significant for tetanus (r = -0.32, n=51, p=0.022) but not for diphtheria (r = -0.21, n = 51, p = 0.14).

**Discussion**

In this study of the humoral immune response in the great snipe, birds responded to vaccination with diphtheria and tetanus toxoid by producing specific antibodies to both antigens used. There was a correlation between the responses to the two antigens and in particular, individuals responding strongly to diphtheria also responded strongly to tetanus. This pattern is very similar to that of other studies that have used these two antigens (Svensson *et al.* 1998; Råberg & Stjernman 2003; Westneat, Hasselquist & Wingfield 2003; Kilpimaa *et al.* 2004).

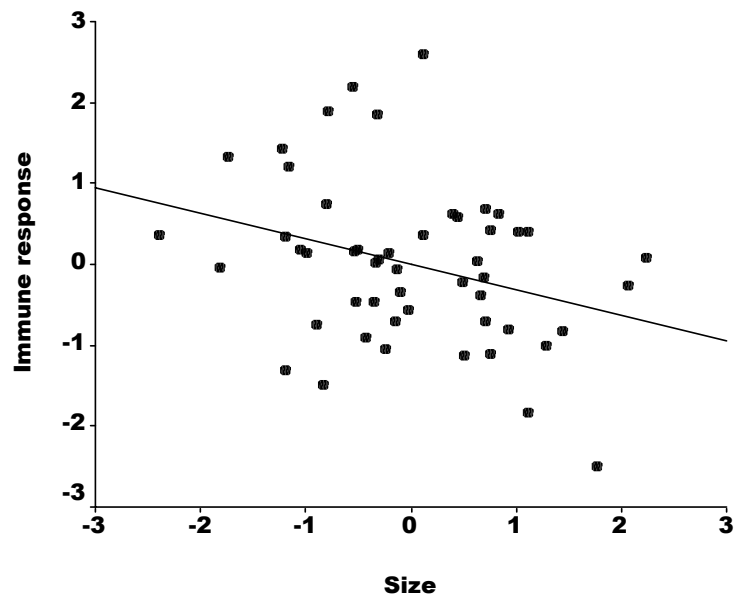

**Fig. 3.** Relationship between size and immune response in great snipe males. There is a weak negative correlation (see text) between these traits suggesting a possible trade-off. The regression line (Y = -0·32X) from the linear regression is included.

A few of the birds were caught and vaccinated in both years of study. The correlation between the antibody response for the two years of these individuals (although not statistically significant due to small sample size) suggest that the strength of the antibody response is stable within individual birds even after one year.

We found no indication that repeated vaccination caused a stronger secondary response. In vertebrates, the general pattern is that secondary antibody responses are much higher than primary responses (e.g., Roitt 1997), and this has also been found in many other studies of wild birds repeatedly exposed to an antigen (Nordling *et al.* 1998; Råberg & Stjernman 2003; Saino *et al.* 2003). However, in all these studies birds were repeatedly injected two or more times with 3-4 weeks between vaccinations, whereas in our study the second injection was conducted about 12 months later. This then may suggest that the time lag between a primary and secondary exposure to an antigen/pathogen may influence the strength of the secondary response. Note, however, that the time interval between injection and collection of a plasma sample to measure the response was approximately 12 days both for the primary and secondary responses. This may influence our data because the timing of the peak antibody titer differs between the primary and secondary response; primary responses peak around two weeks after injection whereas secondary responses peak around 7-9 days after the injection (Hasselquist et al. 1999; Owen-Ashley, Hasselquist & Wingfield in press). Hence, we cannot exclude that we have measured the secondary antibody response in great snipe when it sloped down from the peak, and compared it with peak primary responses.

The strength of the immune response was not affected by the body mass controlled for structural size (a standard measure of condition) of individuals. The effect of individual condition on immunocompetence has been studied in many species and often a positive relationship has been found (see for example Brinkhof et al. 1999; Cichoń 2000; Reid, Arcese & Keller 2003). Effects of condition are often studied by experimentally altering nutritional status or other growing conditions of nestlings. Such studies have often found negative effects on immune response when condition has been suppressed (Lochmiller, Vestey & Boren 1993; Sorci, Soler & Møller 1997; Hõrak et al. 1999; Fargallo et al. 2002; Naguib et al. 2004) and positive effects when food was provisioned (Saino, Calza & Møller 1997; Hoi-Leitner et al. 2001). The amount of antibody production may respond rapidly as condition changes (Cook 1991). In the present study, we found no effect of body condition on humoral immune response in adult great snipe males. One reason for this could be that we did not have large enough sample sizes to detect small effects or that the condition of males during our sampling period may have been influenced by the lekking behaviour. A few published studies have found similar results. For example in great tits, body mass was unrelated to humoral immune response (Snoeijs et al. 2004). Cell mediated immunity in Eurasian treecreepers was not related to habitat quality or brood manipulations (Suorsa et al. 2004), and in house sparrows, birds with a protein poor diet even had a higher humoral immune response than individuals with a high protein diet (Gonzalez et al. 1999). These contrasting results may be explained by the complexity of interactions between condition, nutrition, workload, hormones and immunity, where other factors may cancel out any effects of e.g. condition on immunocompetence.

Other traits that has been used to link condition with immune response include fluctuating asymmetry (Møller & Pomiankowski 1993) and sexual ornaments (Andersson 1994; Brown 1996). Asymmetry has been shown to increase in birds challenged with antigens as chicks (Fair, Hansen & Ricklefs 1999) and in butterflies, symmetry correlates with immune response (Rantala et al. 2000). In contrast to this, we could not find any effect of asymmetry on humoral immune response in this study. Sexual selection has been studied intensely in this species during the last decades and results show that this subject is complicated. The amount of white on the tail feathers have probably been subject to sexual selection but there are no clear evidence that males with more white on the tail are preferred by females in the present situation (Höglund, Eriksson & Lindell 1990; Sæther et al. 2000). We found no relationship between the size of a secondary sexual ornament, i.e. the amount of white on the tail, and antibody responses in the present study, and similar results have been found in red-winged blackbirds (Westneat *et al.* 2003). On the other hand, strong positive correlations between a sexual ornament and humoral immune response have been found in collared flycatchers (Andersson 2001), barn swallows (Saino *et al.* 2003) and in pheasants (Ohlsson *et al.* 2002, 2003). Sexually selected behavioural traits such as song complexity may also be related to the immune response (Duffy & Ball 2002). The relation between ornaments and immune defence seem to be highly dependent on what part of the immune system that is studied (Gonzalez *et al.* 1999; Møller & Petrie 2002; Faivre *et al.* 2003), indicating that these interactions may be very complex. In one study of house sparrows, there was even a negative relationship between male ornamentation and immune response (Møller, Kimball & Erritzoe 1996). Some ornaments also consist of caratenoid pigmentation, substances that may influence immune reactions directly (McGraw & Ardia 2003).

The only positive relationship between humoral immunity and morphology found in our study was a weak negative relationship between antibody production and structural body size. This relationship could reflect a trade-off between investment in immune response and growth (Lochmiller *et al.* 1993; Soler *et al.* 2003, but see Coltman *et al.* 2001). In contrast, there was no effect of size on humoral imunocompetence in two great tit populations (Snoeijs *et al.* 2004). The effect of age on various components of the immune system have recently been studied in various birds and the general trend seems to be that the immune response declines as individuals get older (Cicho´n, Sendecka & Gustafsson 2003; Reid *et al.* 2003; Råberg & Stjernman 2003; Saino *et al.* 2003). However, in agreement with our study, no such effect of age was found in great tits (Snoeijs *et al.* 2004).

We also investigated the effect of MHC genes on the immune response. Råberg et al. (2003) have shown that antibody production to at least one of the antigens used in this study (tetanus) have a strong heritable component (see also Svensson, Sinervo & Comendant 2001). Similar results have also been found for other antigens (Gross *et al.* 1980; Siegel & Gross 1980; Roulin *et al.* 2000). MHC genes are good candidates for mediating this heritable effect since they control which antigens that are able to trigger the immune response (Suri *et al.* 2003). Different aspects of MHC genetics may influence the antibody response. MHC heterozygosity, for example, has been shown to affect T cell function and longevity in mice (Salazar *et al.* 1995). Furthermore MHC heterozygous mice have better resistance to multiple infections (Penn *et al.* 2002), and three-spined sticklebacks with many MHC alleles suffer less from parasitic infection than individuals with few alleles (Wegner *et al.* 2003; Kurtz *et al.* 2004). Certain MHC alleles are also known to enhance immunity to parasitic infections (Clare *et al.* 1985; Bacon 1987; Lamont, Bolin & Cheville 1987) as well as immune response to non-harmful antigens in chicken (Taylor Jr *et al.* 1987; Liu *et al.* 2002; Zhou & Lamont 2003)) and in mammals (Poland 1998; Quinnell *et al.* 2003). But variation in other genes also seems to be important in this respect (Lillehoj *et al.* 1989). Selection for different amounts of antibody response has been shown to affect MHC allele frequencies and disease resistance in chicken (Pinard *et al.* 1993a,b; Dunnington *et al.* 1996). Furthermore, MHC genotype affects susceptibility to bacterial infection in fish (Langefors *et al.* 2001; Lohm *et al.* 2002; Wedekind *et al.* 2004) and in humans both heterozygosity and certain MHC class I alleles influence the outcome of viral infection (Carrington *et al.* 1999; Jeffery *et al.* 2000) as well as many other diseases (reviewed in Thorsby 1997). We could not find any relationship between MHC genetics and strength of humoral immune response in the great snipe. This was true both for MHC heterozygosity, differences in MHC alleles and MHC genotypes. Maybe the response to tetanus and diphtheria antigens is not strong enough to detect any differences, or maybe other genes or ecological factors are more important than MHC class II B genes in determining the strength of the antibody response.

The reason for variation in amount of immune response in vertebrates remains obscure, despite much effort to unravel the importance of different genetic and ecological factors. Many studies show seemingly contradictory results depending of what part of the immune system and what species is being studied. Hence, more observational, correlative studies (like this one) and, in particular, controlled experimental studies are needed in order to understand these complex interactions.

**Acknowledgements**

We thank S. L. Svartaas for field assistance and D. Sejberg, G. Engström and B Rogell for lab assistance. Permissions for capture, ringing, blood sampling and immunisation of birds were given by Stavanger Museum and the Norwegian Animal Research Authority. Financial support were given by Helge Ax:son Johnsons Stiftelse and Stiftelsen för zoologisk forskning (to R.E.), the Swedish Research Council for Environment, Agricultural Science and Spatial Planning (Formas), Carl Tryggers Stiftelse, Crafoordska Stiftelsen and Magn. Bergvalls Stiftelse (to DH), the Research Council of Norway (to SAS) and the Swedich Research Council, VR (to JH).